\documentclass[12pt, letterpaper, epsfig, psfig, subfigure]{article}          
\usepackage{osajnl}

\newcommand\ahwp{WP}

\newcommand{\simlt}{\,\hbox{\lower0.6ex\hbox{$\sim$}\llap{\raise0.6ex\hbox{$<$}}}\,}
\def\beq{\begin{equation}}
\def\eeq{\end{equation}}
\def\beqna{\begin{eqnarray}}
\def\eeqna{\end{eqnarray}}
\long\def\symbolfootnote[#1]#2{\begingroup
\def\thefootnote{\fnsymbol{footnote}}\footnote[#1]{#2}\endgroup}

\begin{document}

\bibliographystyle{unsrt}

\title{Analysis of performance of three- and five-stack achromatic half-wave 
plates at millimeter wavelengths}

\author{Tomotake Matsumura}
\address{School of Physics and Astronomy
\\ University of Minnesota, Twin Cities
\\ 116 Church St. SE, Minneapolis, Minnesota, 55455}
\email{tmatsumu@physics.umn.edu}
\address{Current address: California Institute of Technology
\\ 1200 E. California Blvd. Mail Code 59-33 Pasadena, CA 91125}
\email{tm@caltech.edu}

\author{Shaul Hanany}
\address{School of Physics and Astronomy
\\ University of Minnesota, Twin Cities
\\ 116 Church St. SE, Minneapolis, Minnesota, 55455}
\email{hanany@physics.umn.edu}

\author{Bradley R. Johnson}
\address{Department of Astrophysics
\\ University of Oxford
\\ Keble Road Oxford OX1 3RH England, UK}
\email{bjohnson@physics.ox.ac.uk}

\author{Terry J. Jones}
\address{School of Physics and Astronomy
\\ University of Minnesota, Twin Cities
\\ 116 Church St. SE, Minneapolis, Minnesota, 55455}
\email{tjj@astro.umn.edu}

\author{Prashanth Jonnalagadda}
\address{Department of Computer Science and Engineering
\\ University of Minnesota, Twin Cities
\\ 200 Union Street SE, Minneapolis, Minnesota, 55455}
\email{jonn0006@umn.edu}


\begin{abstract} 

We study the performance of achromatic half-wave plates (AHWP) as a function
of their construction parameters, the detection bandwidth of a
power detector operating in the millimeter wave band, and the spectral 
shape of the incident radiation. We focus particular
attention on the extraction of the degree of incident polarization
and its orientation angle from the intensity measured as a function 
of AHWP rotation angle, which we call the IVA (intensity versus angle). 
We quantify the phase offset of the IVA and point to potential systematic
errors in the extraction of this offset in cases where the incident spectrum is 
not sufficiently well known. We show how the phase offset and modulation 
efficiency of the AHWP depend on the relative angles between the plates
in the stack and find that high modulation efficiency can be achieved with 
alignment accuracy of few degrees. 

\end{abstract}


\maketitle 

\section{Introduction}

Recent experimental efforts in observational cosmology have been
focused on searching for a signature from an inflationary period that
occurred a short instant after the big bang. This signature is
predicted to be imprinted in the polarization of the cosmic microwave
background (CMB) radiation.  Inflation predicts an inflationary
gravitational-wave background (IGB) that left a particular pattern of
linear polarization on the CMB. This pattern is different from that
originating from primordial density anisotropy, which is the main
source for the spatial intensity fluctuations over the sky and for a
different, stronger linear polarization pattern. In the past few years,
several groups have started to characterize the polarization
signal coming from the primordial density 
anisotropy~\cite{kovac02,page06}. The polarization pattern from the
IGB is expected to be at least an order of magnitude smaller 
and it has not been detected yet. 

Thorough understanding of foregrounds and good control of systematic
errors will be required to extract the small signal from the IGB. Both
requirements lead to polarimeter designs that implement broad
frequency coverage. Examples of such polarimeters are 
EBEX, a NASA supported balloon-borne experiment~\cite{oxley04} that 
is being constructed by the authors of this paper and their collaborators, 
BICEP/SPUD, CLOVER, QUIET, SPIDER, PAPPA, and PolarBear 
\cite{kovac07,taylor04,quiet_url,montroy06,kogut06,polarbear_url}.

A common technique to measure linearly polarized radiation 
is to use a rotating half-wave plate
(HWP) together with a linear polarizer. The technique has been 
used extensively in the optical and IR wavelengths~\cite{jones88,platt91,leach91,murray97}. 
The first experiment to report CMB polarization results with this technique
was MAXIPOL~\cite{johnson06,wu06}. 
Although a HWP is a device that operates over narrow band 
of frequencies there are standard techniques to construct an
'achromatic HWP' (AHWP) that operates over a much broader 
range~\cite{pancharatnam55,title75,title81,tinbergenbook,hanany05}. 
An AHWP is a stack of birefringent plates 
that are aligned with specific relative orientation angles between 
their optic axes. With an appropriate choice of angles 
it is possible to achieve modulation efficiency that 
is close to 100~\% over a large fraction of the
millimeter wavelength band~\cite{hanany05}. 

The linear polarization content of incident radiation can be characterized 
in terms of two parameters, the degree of polarization $P_{in}$ and the 
orientation angle $\alpha_{in}$. An equivalent set is the normalized Stokes parameters
$Q_{in}/I_{in}$ and $U_{in}/I_{in}$ (see Section~\ref{sec:model} for a definition). 
To reconstruct these parameters from the signal detected
by the instrument it is essential to quantify the extent to which 
the polarimeter itself changes the input parameters. 

In this paper, we present a computational study of the effects
introduced by three- or five-stack AHWPs that are designed to 
fit CMB polarimeters operating in the range $120-480$~GHz. This work 
is motivated by EBEX
that will implement a rotating 5-stack AHWP to search for the faint
signals from the IGB. In Sections~\ref{sec:model} and~\ref{sec:figureofmerit}
we describe the mathematical formalism and define the figures of 
merit that are used to reconstruct the
state of incident polarized light from the measured intensity. 
In Section~\ref{sec:result} we use the figures of merit to quantify 
how well the incident polarization can be reconstructed. 
Section~\ref{sec:spectrum} 
discusses effects that arise from uncertainties in the spectrum
of the incident radiation. In Section \ref{sec:performance.vs.theta}
we assess the performance of
an AHWP as a function of its construction parameters. 
A summary of the key conclusions is given in Section~\ref{sec:discussion}.

\begin{figure}[t] 
\centering
\includegraphics[width=15.cm, angle=180]{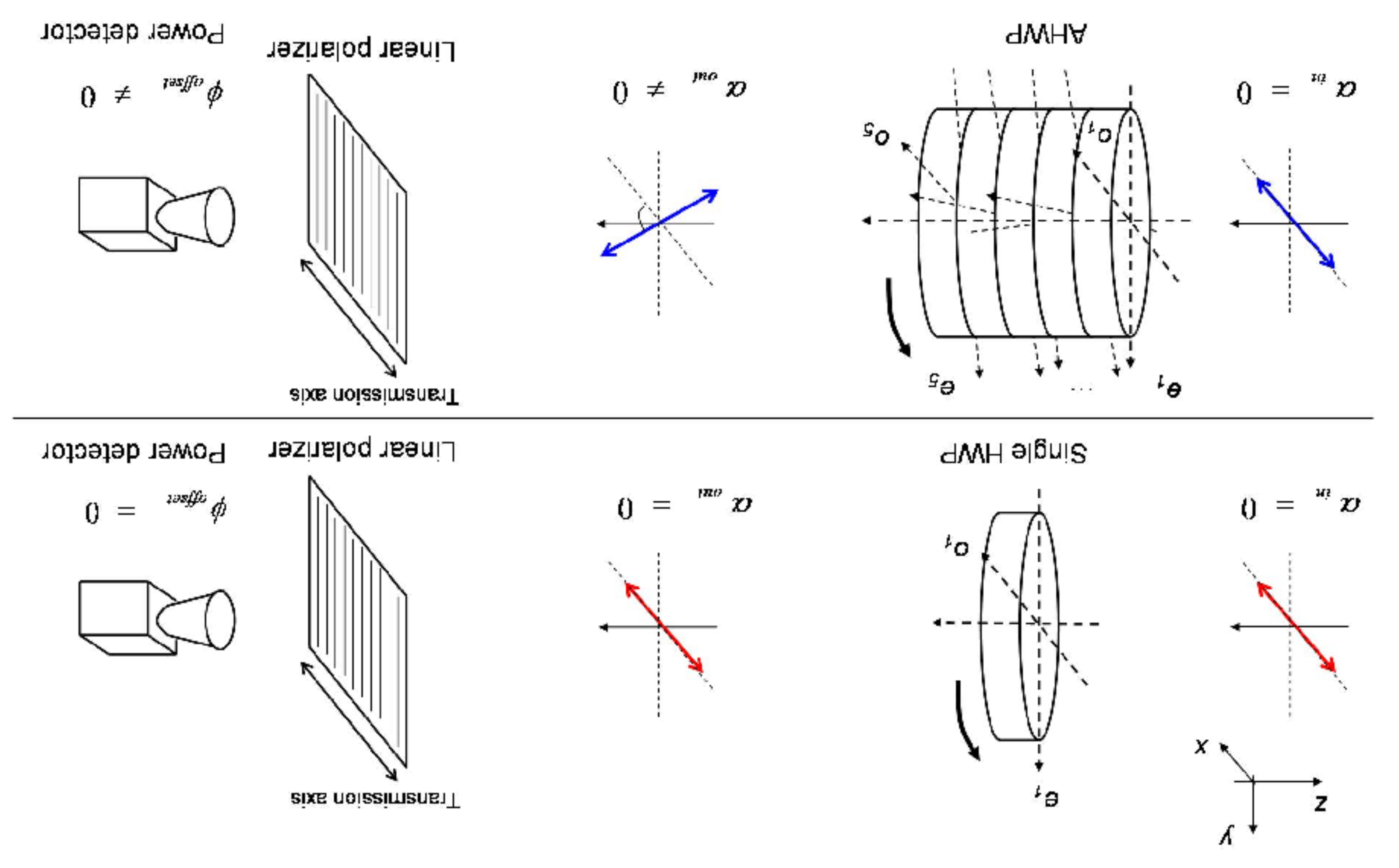} \vspace{-1.cm}
\caption{\footnotesize \setlength{\baselineskip}{0.95\baselineskip} 
	A schematic diagram of the HWP polarimeter model. 
	The transmission axis of a linear polarizer is parallel to the $x$ axis. }
\label{fig:setup}
\end{figure}


\section{Polarimeter Model}
\label{sec:model}

We consider a polarimeter that consists of an AHWP that rotates at  a 
frequency $f_{0}$, a linear polarizer, and a power detector (e.g. a
bolometer), as shown in Figure~\ref{fig:setup}. 
The angle $\rho$ gives the rotation angle of the AHWP around its axis
of symmetry $z$.
In such a polarimeter information about
the incident polarization is contained in the intensity that is
detected by the detector as a function of $\rho$. To a good
approximation the detected intensity is sinusoidal as a function of
$\rho$ with a frequency of $4 f_0$ when there is a high
signal-to-noise ratio.  Our primary interest in this paper is to
analyze the detected intensity as a function of $\rho$, which we call
IVA (intensity vs. angle), with the purpose of reconstructing the
incident polarization.

We use Mueller matrices to describe the signal in the approximation of
normal incidence on the \ahwp. We neglect the effect of absorption 
by the wave plate or effects of reflections between media that have 
different indices of refraction. 

Consider an input Stokes vector $\vec{S}_{in}$ of radiation
propagating along the $z$ axis that is incident on the
polarimeter. The  Stokes vector incident on the detector, which we call 
the output Stokes vector, is
\begin{equation}
\vec{S}_{out} = G \prod_{i=1}^{m} [ R(-\rho-\theta_i) \Gamma_{i}( \Delta\delta) 
  R(\rho+\theta_i)] \vec{S}_{in}(\alpha_{in}, P_{in}), 
\label{eq:Stokesout}
\end{equation}
where
\beqna
\Delta\delta &=& 2\pi \frac{\nu}{c} |n_o-n_e| d , \\ 
\label{eq:Deldel}
\vec{S}_{in} &=&  (I_{in}, Q_{in}, U_{in}, 0) \nonumber \\
&=& I(\nu) (1, P_{in} \cos{2\alpha_{in}}, P_{in} \sin{2\alpha_{in}}, 0),
\label{eq:Stokesin}
\eeqna
$G$ is the Mueller matrix of the linear polarizer, $R$ is a rotation
matrix, $\Gamma$ is the Mueller matrix of a retarder, and
$\vec{S}_{in}$ is the Stokes vector of the incident radiation, which
is a function of the polarization angle $\alpha_{in}$ and of the
degree of polarization $P_{in}$.   Information about 
the spectrum of the incident radiation is contained in $I(\nu)$. 
We initially assume that the intensity of the incident radiation is
constant with frequency, $I(\nu) = I_{0} = const$ .  
We discuss the effects of a non-constant 
incident spectrum in Section~\ref{sec:spectrum}.
Equation~\ref{eq:Stokesout} assumes $m$ wave plates in the stack; in
this paper $m=1, 3$ or $5$. The variable $\Delta\delta$ is the
retardance of a single wave plate and is a function of the ordinary
and extraordinary indices of refraction $n_o$ and $n_e$, respectively,
the thickness of a single wave plate $d$, and the electromagnetic frequency
of light $\nu$.  Throughout this paper, we assume that the incident radiation
is not circularly polarized. We also assume that $\alpha_{in}$ and $P_{in}$
are independent of $\nu$. The components of the Mueller matrices are
\begin{eqnarray}
\label{eq:retard}
\Gamma(\Delta \delta) =
\left [\begin{array}{cccc}
1 & 0 &                                   0 & 0 \\ 
0 & 1 &                                   0 & 0 \\
0 & 0 & \cos{\Delta \delta} & -\sin{\Delta \delta} \\
0 & 0 &   \sin{\Delta \delta} & \cos{\Delta \delta} \\
\end{array}\right ],
\end{eqnarray}
\begin{eqnarray}
R(\theta) =
\left [\begin{array}{cccc}
1 &                     0 &                      0 & 0 \\ 
0 & \cos{2\theta} & -\sin{2\theta} & 0 \\
0 &  \sin{2\theta} & \cos{2\theta} & 0 \\
0 &                     0 &                      0 & 1
\end{array}\right ],
\end{eqnarray}
and
\begin{eqnarray}
G = \frac{1}{2}
\left [\begin{array}{cccc}
1 & 1 & 0 & 0 \\ 
1 & 1 & 0 & 0 \\ 
0 & 0 & 0 & 0 \\ 
0 & 0 & 0 & 0 
\end{array}\right ]. 
\end{eqnarray}
As shown in Figure~\ref{fig:setup}, we choose the transmission axis of 
the ideal linear polarizer to be aligned with the $+x$ axis. 
We define all the angles of rotation about the $z$ axis 
with respect to the transmission axis of the grid.
According to the usual convention, angles increase in the counter-clockwise direction
from the $+x$ axis in the $xy$ plane. The relative orientation of plate
$i=2,3...$ in the stack relative to the first plate is given by
$\theta_i$. The ordinary axis of the first plate is aligned with the
$x$ axis when $\rho=0$~degrees. We use the notation $\vec{\theta}$ 
to denote the entire set of relative orientation angles. 

The output of the detector is a function of its detection 
bandwidth and the intensity term of $\vec{S}_{out}$.
We assume that the detector has
top-hat response of width $\Delta \nu$ about a center frequency
$\nu_c$.  Note that the limit $\Delta \nu \rightarrow 0$ is equivalent
to the case of illuminating the polarimeter with monochromatic
light. With these assumptions the first element of the output Stokes
vector can be written as
\begin{equation}
\langle I_{out} \rangle(\nu_c, \Delta \nu, \alpha_{in}, P_{in}, \vec{\theta}, \rho) 
 = \int_{\nu_{c}-\frac{\Delta \nu}{2}}^{\nu_{c}+\frac{\Delta \nu}{2}} 
    I_{out} (\nu, \alpha_{in}, P_{in}, \vec{\theta}, \rho)\mbox{d}\nu.
\label{eq:Maluslaw_freqsum}
\end{equation}
A plot of $\langle I_{out} \rangle$ as a function of $\rho$ is the IVA. 
(Throughout this paper angle brackets $\langle\,\rangle$ denote integration over frequency.)
For a single HWP $I_{out}$ of Equation~\ref{eq:Maluslaw_freqsum} is
\begin{equation}
I_{out}(\nu) = \frac{I_{0}}{2}(1+P_{in}\cos{2\alpha_{in}}\cos^2{\frac{\Delta\delta(\nu)}{2}}
+ P_{in}\sin^2{\frac{\Delta\delta(\nu)}{2}} \cos{(4\rho-2\alpha_{in})}) 
\label{eq:singleHWPIVA_analytical}
\end{equation}
and an analytic integration over any bandwidth is straight forward. For a three- 
and five-stack AHWP the analytic expressions are more complicated. 

It is the goal of this paper to discuss quantitatively how the
amplitude and phase of the IVA depend on the construction parameters
of the AHWP. Specifically, we make a quantitative mapping between 
the measured amplitude and phase of the IVA and the 
two parameters characterizing the incident polarization, the degree
of input polarization $P_{in}$ and its orientation angle $\alpha_{in}$. 

When there is a finite detection bandwidth $\Delta \nu \ne 0$ the
amplitude and phase of the IVA are calculated in the following way.
We calculate the intensities as a function of angle $\rho$ for each
frequency within the bandwidth. We then sum the calculated intensities
angle by angle to obtain a final IVA. The amplitude and phase are
determined from that IVA. 

All of the analysis in this paper is
computational. IVAs have been calculated as a function of
various parameters of the incident radiation and of the construction
of the HWP. Many of the results were calculated by two independent
computer codes to check for errors. Where practical, the results
were compared to analytical calculations and agreement has been
verified. 

\section{Figures of Merit}
\label{sec:figureofmerit}

\begin{figure}[t]
\centering
\includegraphics[width=12.5cm, angle=270]{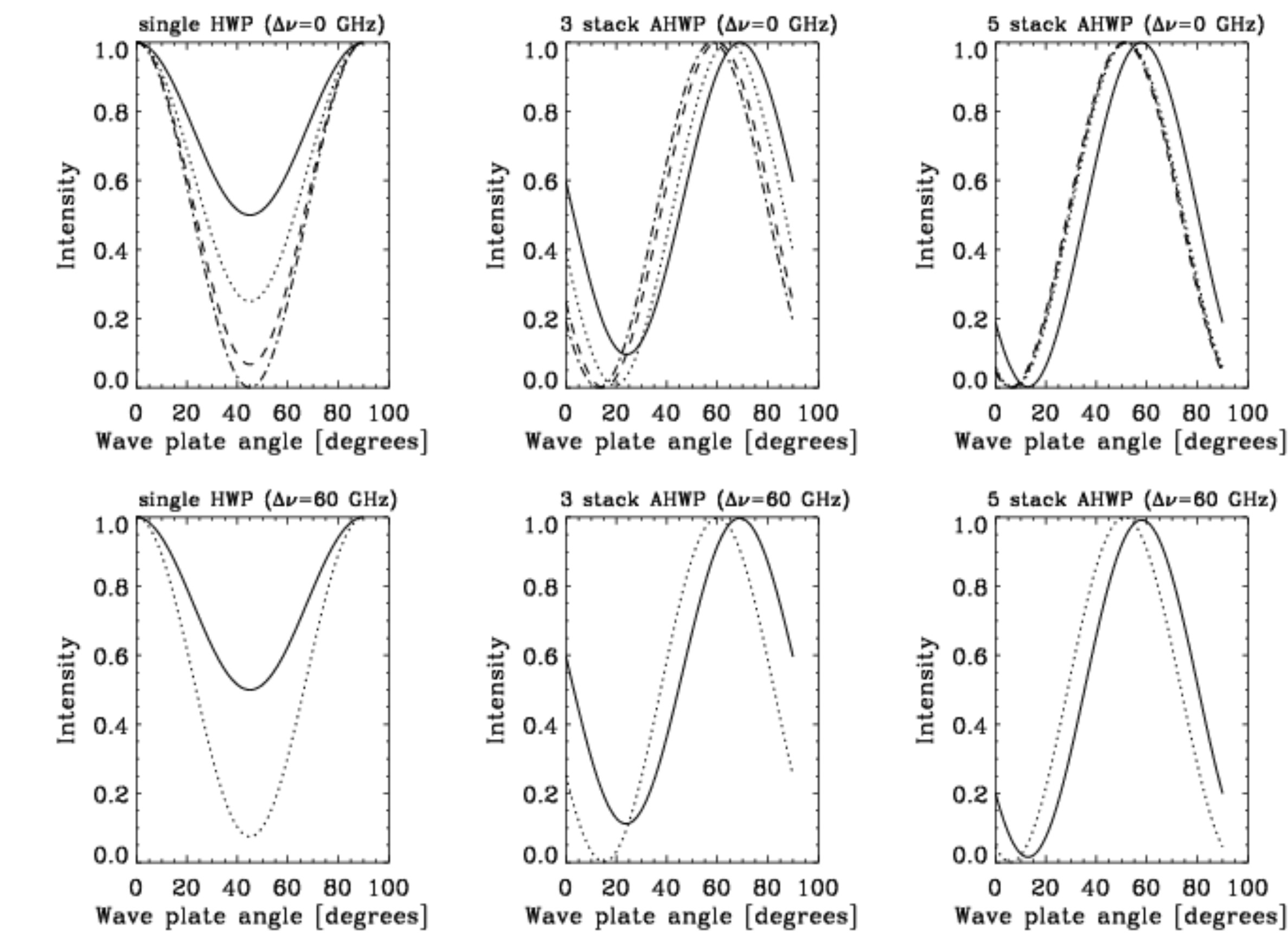}
\vspace{-1cm}
\caption{\footnotesize \setlength{\baselineskip}{0.95\baselineskip}
	IVA for monochromatic light (top panels) and for broadband
	radiation (bottom panels) for a single HWP, a three-stack AHWP,
	and a five-stack AHWP (left to right). See
	Table~\ref{tab:pars} for the parameters of the plates and
      for the details about the simulations used for the calculations. 
      Frequencies of 150~(solid), 200~(dash), 250~(dot), 300~(dash-dot)~GHz are
	used for the case of monochromatic light. For the broadband
	case we use $150\pm30$~GHz~(solid) and $250\pm30$~GHz~(dot).
	In all the panels, the maximum intensity is normalized to 1.}
\label{fig:IVA} 
\end{figure}

Figure~\ref{fig:IVA} shows the IVA for the case of a single sapphire
HWP and for an AHWP made of a stack of three and five sapphire
plates. Table~\ref{tab:pars} lists the parameters of the plates used
to generate these IVAs. The thickness of each wave plate gives $\Delta
\delta = \pi$ when $\nu_{WP} = 300$~GHz.  The top panels in
Figure~\ref{fig:IVA} show the IVA for a monochromatic detection
bandwidth ($\Delta \nu = 0$).  In the panels on the bottom the
detection bandwidth is $\Delta\nu=60$~GHz. The incident light is
polarized parallel to the transmission axis of the grid, $\alpha_{in}
= 0$~degrees.

\begin{table}[t]
\begin{center}
\begin{tabular}[t]{|c|c|} \hline
Incident intensity & $I = 1$\\ \hline
Indices of refraction of sapphire\cite{loewenstein73}& $n_o=3.047, n_e = 3.364$ \\ \hline
Thickness of each wave plate, $d$  & 1.58 mm  ($\leftrightarrow \nu_{WP}$ = 300 GHz) \\ \hline
Bandwidth of frequency, $\nu_c + \Delta\nu$ & 150~$\pm$~30~GHz, 250~$\pm$~30~GHz\\ \hline
Orientation angles of three-stack AHWP, $\vec{\theta}_3$ & (0, 58, 0)  degrees  \\ \hline
Orientation angles of five-stack AHWP, $\vec{\theta}_5$  & (0, 29, 94.5, 29, 2) degrees \\ \hline
Resolution of frequency                             & 0.5 GHz         \\ \hline
Resolution of wave plate angle                      & 0.1 degrees           \\ \hline
\end{tabular}
\end{center}
\vspace{-1.cm}
\caption{\footnotesize \setlength{\baselineskip}{0.95\baselineskip}
	 Parameters of the wave plates and parameters used 
      in the simulations to calculate the IVA. } 
\label{tab:pars}
\end{table}

\begin{figure}[t]
\centering
\includegraphics[width=17cm, angle=180]{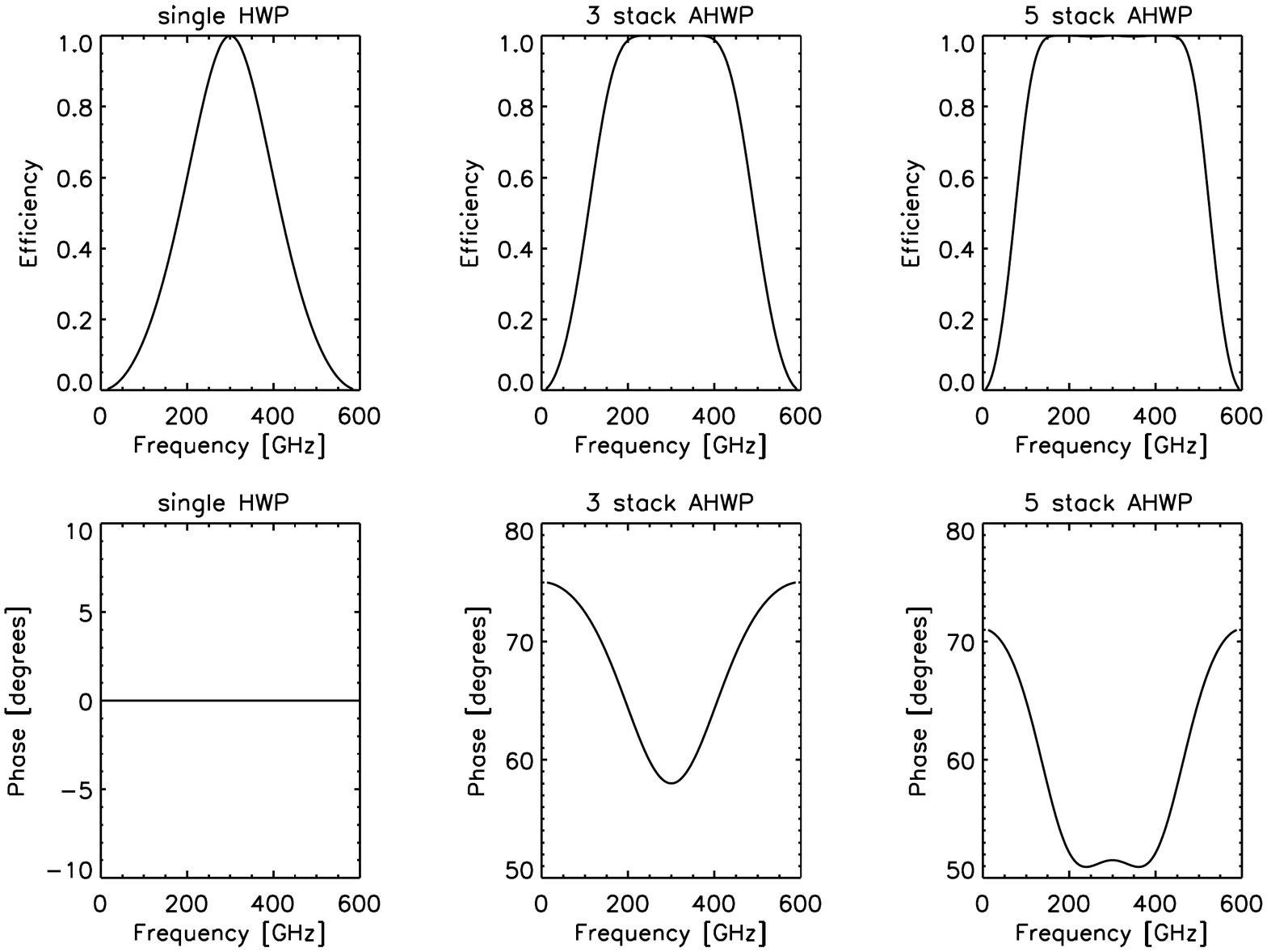}
\vspace{-1.cm}
\caption{\footnotesize \setlength{\baselineskip}{0.95\baselineskip} 
	Modulation efficiency $\epsilon = \epsilon(\nu, \Delta\nu=0,
	\alpha_{in}=0,\vec{\theta})$ (top) and the phase offset
	$\phi_0 = \phi(\alpha_{in}=0, \nu, \Delta\nu=0, \vec{\theta})$ (bottom)
	for the single HWP (left) and the three- (middle) and the
	five-stack (right)  as a function of frequency.     }
\label{fig:fig_eff_phase.ps}
\end{figure}

Several generic features are apparent.  The reduction in the amplitude
of the IVA with a single HWP (left column) is a consequence of its
chromaticity.  Linear input polarization becomes     elliptical when
it passes through a wave plate that is optimized for a different
frequency. There is substantially smaller reduction in amplitude of
modulation for the 3-stack (middle column) or 5-stack (right column). 
However, whereas  for a single plate   the phase of the IVA is
the same between different frequencies, or with a broad detection
bandwidth,   it becomes   a function of
frequency for the case of an AHWP.  We define the phase angle $\phi$ of the
IVA as
\begin{equation}
\langle I_{out} \rangle = A_{0} + A_{4} \cos{(4\rho-4\phi)},
\label{eq:IVA}
\end{equation}
where $A_{0}$ and $A_{4} $ denote the average level and 
the modulation amplitude of the IVA, respectively. For the case of a single HWP the 
forms of $A_{0}$ and $A_{4}$ are
\begin{equation}
A_{0} = \frac{I_{0}}{2} \left( \Delta \nu + P_{in}\cos{2\alpha_{in}} \int_{\nu_{c}-\frac{\Delta \nu}{2}}^{\nu_{c}+\frac{\Delta \nu}{2}}\cos^2{\frac{\Delta\delta(\nu)}{2}}\, d\nu \right),
\label{eq:DC}
\end{equation}
and 
\begin{equation}
A_{4} = \frac{I_{0} P_{in}}{2} \int_{\nu_{c}-\frac{\Delta \nu}{2}}^{\nu_{c}+\frac{\Delta \nu}{2}}\sin^2{\frac{\Delta\delta(\nu)}{2}}\, d\nu.  
\label{eq:AMP}
\end{equation}
From a comparison of Equations~\ref{eq:singleHWPIVA_analytical}
and~\ref{eq:IVA} it is evident that in this case $\phi = \alpha_{in}/2$ and that it is    
independent of frequency. Since we chose $\alpha_{in}=0$~degrees for
the simulation shown in the left panels of Figure~\ref{fig:IVA}, $\phi=0$~degrees.
However, for an AHWP the phase $\phi$ is a function of the thickness 
of the HWP, the detection bandwidth and the relative orientation
angles. Mathematically 
$\phi = \phi(\nu_c, \Delta \nu, \alpha_{in}, \vec{\theta})$, 
and therefore the IVAs in the middle and right columns of 
Figure~\ref{fig:IVA} show non-zero phase angles. 
We define this overall 'phase offset' of the 3- and 5-stack AHWPs
as $\phi_0$. The quantity $\phi_0$ is the value of $\phi$ when $\alpha_{in} = 0$~degrees
(e.g. $\phi_0 \sim 65$~degrees for the solid line of the middle bottom panel of 
Figure~\ref{fig:IVA}).
\begin{figure}[t]
\centering
\includegraphics[width=10cm, angle=270]{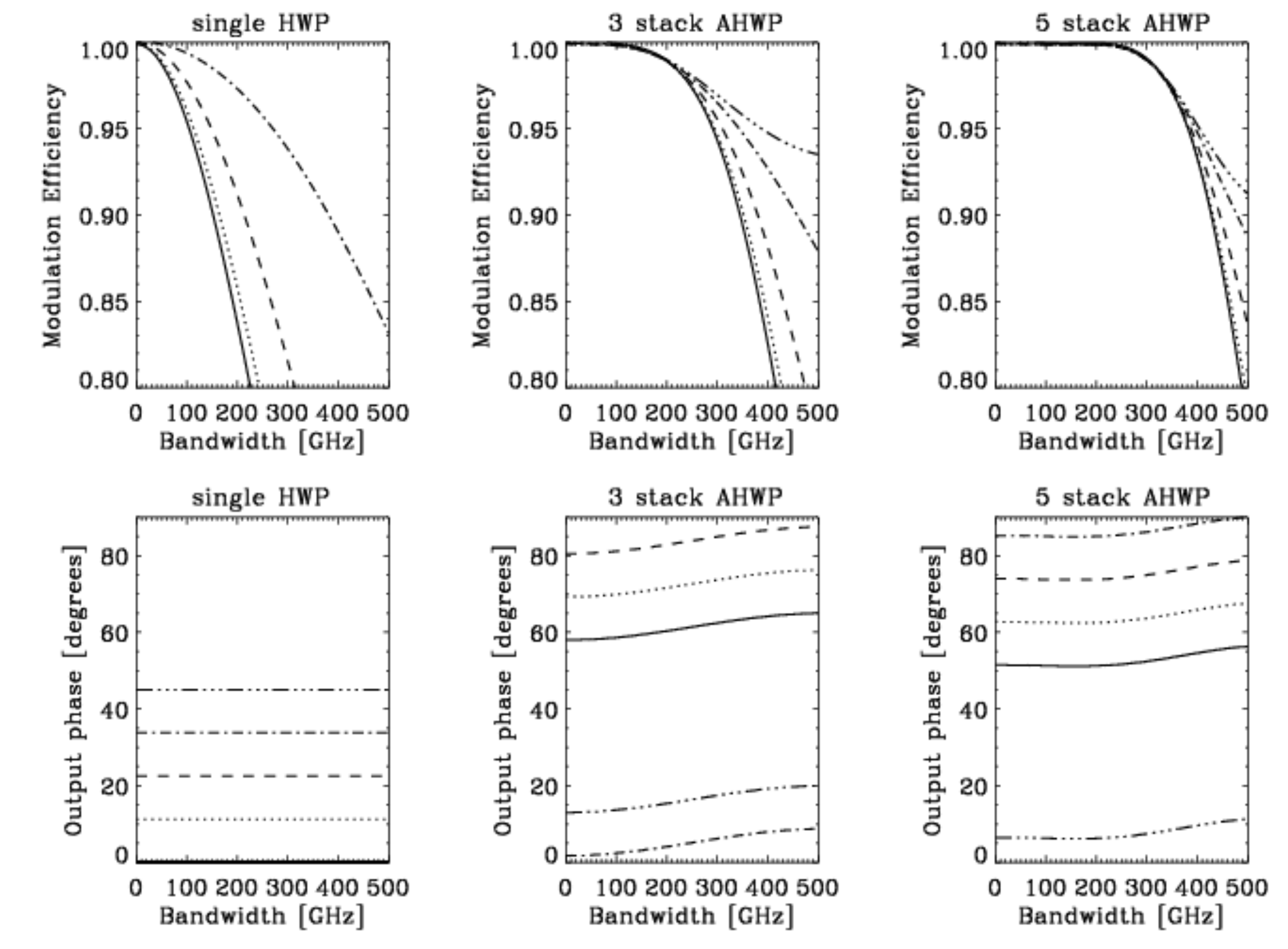}
\vspace{-1.cm}
\caption{\footnotesize \setlength{\baselineskip}{0.95\baselineskip} 
	 Top: Modulation efficiency of the single HWP, the three- and
	 the five-stack AHWPs  as a function of detection bandwidth
                          for   input polarization
	 angle of 0 (solid line), 22.5 (dot), 45 (dash), 67.5
	 (dot-dash), and 90 (three-dot dash) degrees. Bottom: 
	 Output phase angle of the single, three-, and five-stack 
	 as a function of detection bandwidth for  
	 the same input polarization angles as the top panels. For
	 both the modulation efficiency and the phase, $\nu_c =
	 \nu_{WP}$.}
\label{fig:me_bandwidth_alpha.ps}
\end{figure}

A useful figure of merit for the operation of a polarimeter is the
'modulation efficiency'~\cite{tinbergenbook} defined as 
\begin{equation}
\label{eq:me}
\epsilon = \epsilon(\nu_c, \Delta \nu, \alpha_{in}, P_{in}, \vec{\theta}) 
       =  \frac{P_{out}} {P_{in}}. 
\end{equation}
The efficiency $\epsilon$ 
is a measure of the de-polarization introduced by the polarimeter
and is an essential element in reconstructing the incident 
polarization $P_{in}$ from the one measured by the experiment $P_{out}$. 
In this paper we calculate $\epsilon$ by extracting 
$P_{out}$ from the IVA. $P_{out}$ is calculated from the ratio,
\begin{equation}
\label{eq:Pout}
P_{out} = P_{out}(\nu_c, \Delta \nu, \alpha_{in}, P_{in}, \vec{\theta})
      = \frac{\langle I_{out} \rangle_{max} - \langle I_{out} \rangle_{min}}
        {\langle I_{out} \rangle_{max} + \langle I_{out} \rangle_{min}}.   
\end{equation}
Here $\langle I_{out} \rangle_{max}$ and $ \langle I_{out} \rangle_{min} $ 
are the maximum and minimum of the IVA for angles $0\leq\rho<90$.
(The modulation efficiency that was calculated in our earlier publication~\cite{hanany05}
assumed a somewhat different functional form for $P_{out}$. See
Section~\ref{sec:performance.vs.theta} for more details.) 
Using Equations~\ref{eq:singleHWPIVA_analytical}, \ref{eq:DC} and~\ref{eq:AMP} it is
straight forward to show that for a single HWP {\it and a single frequency}
\begin{equation}
\label{eq:pout_single}
P_{out}(\nu) = \frac{P_{in}\sin^2{\frac{\Delta\delta(\nu)}{2}}}
              {1 + P_{in}\cos{2\alpha_{in}} \cos^2{\frac{\Delta\delta(\nu)}{2}} }, 
\end{equation}
and therefore
\begin{equation}
\label{eq:eps_single}
\epsilon(\nu) = \frac{P_{out}}{P_{in}} = \frac{\sin^2{\frac{\Delta\delta(\nu)}{2}}}
              {1 + P_{in}\cos{2\alpha_{in}} \cos^2{\frac{\Delta\delta(\nu)}{2}} }.
\end{equation}
There are two cases for which this expression is particularly useful, (i) when $P_{in}$ is 
sufficiently small such that the denominator is approximately 1, 
and (ii) when $\alpha_{in}=45$~degrees (for any level of $P_{in}$). In both of these cases 
\begin{equation}
\label{eq:eps_single_smallP}
\epsilon(\nu) = \sin^2{ \frac{\Delta\delta(\nu)}{2} },
\end{equation}
$\epsilon$ is only a function of the retardance of the HWP and is independent of $P_{in}$.
In the first case $\epsilon$ also does not depend on $\alpha_{in}$. 

We note that instead of using Equation~\ref{eq:Pout} a more generally 
appropriate process for extracting
the modulation efficiency is by fitting the IVA to a harmonic series of 
sine waves and then calculating
\begin{equation}
\label{eq:eps_fourier}
\epsilon = \frac{A_{4}}{A_{0}},
\end{equation}
where $A_{0}$ and $A_{4}$ are the coefficients of the zeroth and fourth harmonic
terms, respectively. For the results presented in this 
paper there is no difference between the two processes. 

The upper panels in Figure~\ref{fig:fig_eff_phase.ps} give the 
modulation efficiency of a single HWP and
of a three- and five-stack AHWP as a function of frequency.
To calculate these efficiencies we analyzed monochromatic IVAs. 
The reduction in efficiency for the three-stack at a frequency of, for
example, 150~GHz can be mapped to the smaller amplitude IVA in
Figure~\ref{fig:IVA} for the same frequency. The set of three
panels shows that a larger number of plates in the stack gives
a broader bandwidth of high modulation efficiency.

In order to reconstruct the polarization angle $\alpha_{in}$  of the
incident polarization it is essential to examine the phase angle of the
IVA.  The lower panels in Figure~\ref{fig:fig_eff_phase.ps} show
$\phi_0$ as a function of frequency. They were also extracted from
IVAs calculated with monochromatic detection bandwidths.  The phase offset
varies with frequency even  near the center frequency $\nu_{WP}$. 

To gain some insight into the effects that we quantify
in subsequent sections let us compare the middle panels of
Figures 2 and 3 in more detail. The solid lines shown in 
Figure 2 were calculated for a center frequency of 150~GHz.
The upper panel in Figure 2 shows the corresponding
monochromatic IVA. The modulation efficiency of $\sim$0.95
and phase angle of $\sim$70~degrees of that IVA can be read
directly from the middle panels of Figures 3 at a frequency of 
$150$~GHz. The lower panel of Figure 2 shows an IVA that has been 
calculated for a detection bandwidth of $\pm~$30~GHz around 
$150~$GHz. It was calculated by  averaging the intensities
at different frequencies. We used a frequency resolution of $0.5~$GHz
(see Table~\ref{tab:pars}). Each of these IVAs has a modulation 
efficiency and 
phase offset that can be read off from the middle panels of 
Figure 3. Both the modulation efficiency and the phase offset 
vary over the bandwidth. As a consequence, the resulting
IVA is a superposition of sine waves with different amplitudes and
phases. Hence the final IVA is also a sine wave, but its amplitude
and phase depends on averaging intensities over frequencies. 

In the next Section we discuss how the efficiency $\epsilon$ and 
the phase angle $\phi$ depend on the construction parameters of the 
stack of HWPs, on the center frequency, and on the 
detection bandwidth, and how to relate them to the parameters of the 
incident polarization $P_{in}$ and $\alpha_{in}$.

\section{Reconstruction of $P_{in}$ and $\alpha_{in}$}
\label{sec:result}

\subsection{Modulation Efficiency and Phase}
\label{subsec:me}

The top panels of Figure~\ref{fig:me_bandwidth_alpha.ps} show the
modulation efficiency of a single HWP, three-stack, and five-stack
AHWPs as a function of bandwidth around $\nu_c = 300$~GHz. The
different curves correspond to different input polarization angles
$\alpha_{in}$. To calculate the modulation efficiency we used
$P_{in} = 1$. For $\alpha_{in} = 0$~degrees modulation efficiency
that is larger than 0.99 is achieved with a bandwidth of 200 (300)~GHz
for the three (five)-stack AHWP, while a single HWP achieves a
bandwidth of only 50 GHz.  
For a given bandwidth the 
modulation efficiency is a function of the orientation 
of the incident polarization $\alpha_{in}$.  So in order to 
reconstruct $P_{in}$, information about $\alpha_{in}$ needs 
to be extracted first from the measured phase angle $\phi$. 

The bottom panels of Figure~\ref{fig:me_bandwidth_alpha.ps} show the
output phase angle $\phi$ as a function of bandwidth around $\nu_c =
300$~GHz. The phase of the single HWP shows flat response over the
bandwidth. The phases of the IVA of the three- and five-stack AHWPs
are a function of bandwidth, a result consistent with the bottom row
of Figure~\ref{fig:fig_eff_phase.ps}. For a given bandwidth the phase angle
$\phi$ has an overall offset $\phi_0$.

The conclusions so far are that if the incident radiation
is known to be fully polarized and the detection bandwidth 
is known, then the orientation angle of the incident polarization 
and the modulation efficiency can be extracted. Alternatively, if the 
orientation angle of the incident fully polarized radiation is known
then the modulation efficiency and an equivalent detection
bandwidth can be extracted. These situations are encountered
in the laboratory when calibrating the polarimeter.

\begin{figure}[t] 
\centering
\includegraphics[width=10.5cm, angle=270]{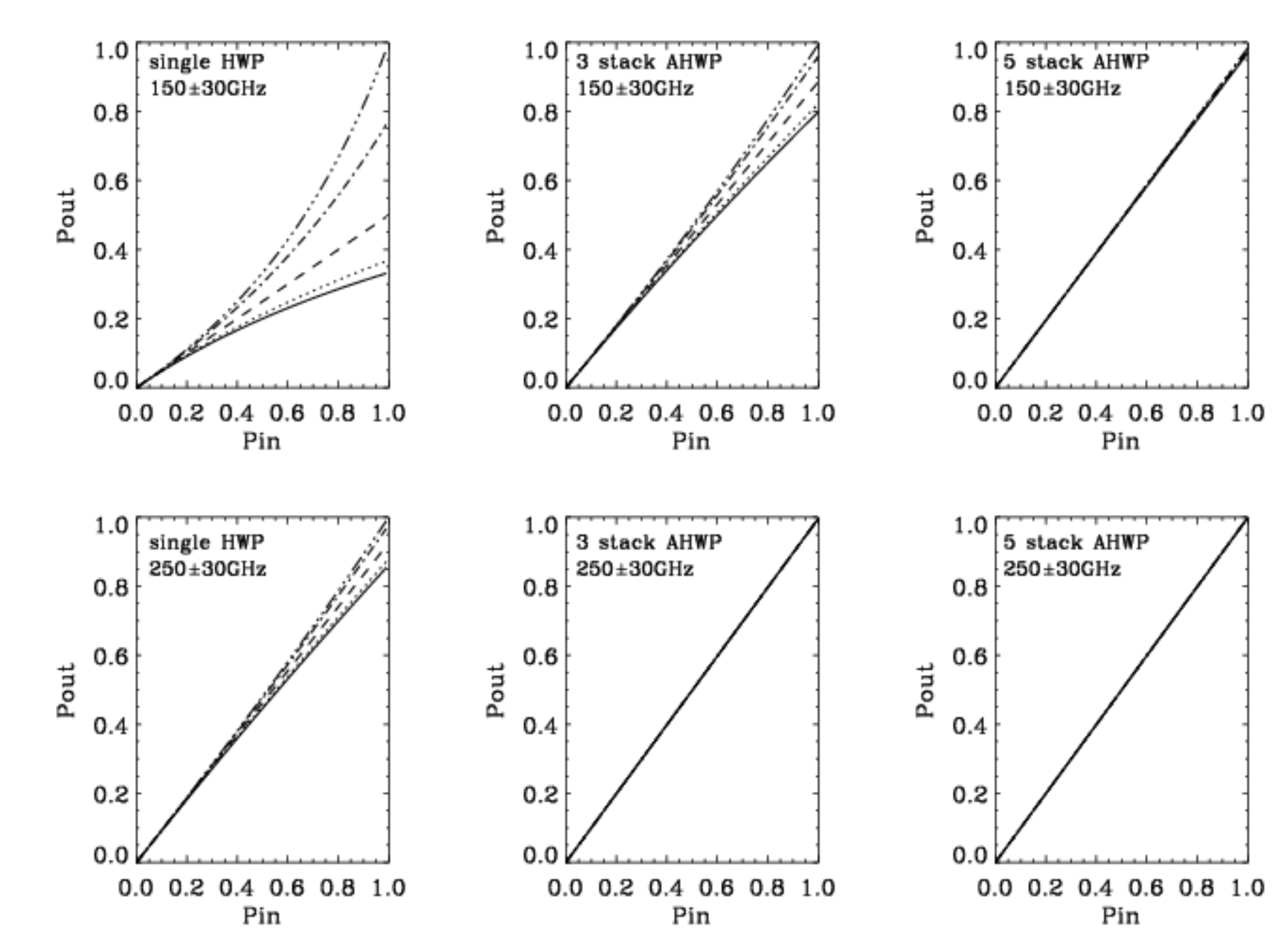}
\vspace{-1.cm}
\caption{\footnotesize \setlength{\baselineskip}{0.95\baselineskip} 
	The extracted degree of polarization $P_{out}$ as a
	function of the degree of polarization of the incident light $P_{in}$
	for the single-, three-, and five-stack. Each curve
	corresponds to the input polarization angle of 0 (solid
	line), 22.5 (dot), 45 (dash), 67.5 (dot-dash), and 90 (three-dot dash)
	degrees.
	The frequency and the bandwidth are $\nu_c\pm\Delta\nu = 150\pm30$~GHz (top)
	and $250\pm30$~GHz (bottom). For all the panels, $\nu_{WP} = 300$~GHz.}
\label{fig:fig_eff_Pin.ps} 
\end{figure} 

\begin{table}[th]
\begin{center}
\begin{tabular}[h]{|c|c|c|} \hline
$\epsilon_{45}  \mbox{}^{+\epsilon_{max}}_{-\epsilon_{min}} $  & 150~$\pm$~30~GHz & 250~$\pm$~30~GHz   \\ \hline
single HWP               &   $0.50^{+0.055}_{-0.045}$        & $0.93^{+0.015}_{-0.015}$            \\ \hline
three-stack AHWP  &   $0.89^{+0.02}_{-0.02}$        & $0.996^{+0.001}_{-0.000}$            \\ \hline
five-stack AHWP     &   $0.976^{+0.002}_{-0.001}$        & $0.999^{+0.001}_{-0.000}$            \\ \hline
\end{tabular}
\end{center}
\vspace{-1.cm}
\caption{\footnotesize \setlength{\baselineskip}{0.95\baselineskip} 
	The modulation efficiency at $P_{in} = 0.1$ with $\alpha_{in}
	= 45$~degrees is shown. The modulation efficiency is
	calculated as a slope of $P_{out}-P_{in}$ relationship in
	Figure~\ref{fig:fig_eff_Pin.ps}. The quoted errors are
	$\epsilon_{max} - \epsilon_{45}$ and $\epsilon_{min} - \epsilon_{45}$, 
	where $\epsilon_{45}$ corresponds to the
	modulation efficiency at $\alpha_{in} = 45$~degrees at $P_{in} = 0.1$. The
	maximum and the minimum modulation efficiency corresponds to
	$\alpha_{in} = 90$ and $0$~degrees, respectively.}
\label{tab:Pout_Pin}
\end{table}

\subsection{$P_{out}$ vs. $P_{in}$}
\label{sec:pout_vs_pin}

The results in Figure~\ref{fig:me_bandwidth_alpha.ps} were calculated
with the assumption of incident polarization $P_{in}=1$. We relax
this assumption in Figure~\ref{fig:fig_eff_Pin.ps}, which shows
$P_{out}$ as a function of $P_{in}$ for various angles $\alpha_{in}$.  The
local slope of each curve is the modulation efficiency $\epsilon$.
The modulation efficiency is a function of both
$\alpha_{in}$ and $P_{in}$. That this is the case for a single HWP is 
evident from Equation~\ref{eq:eps_single}. 
In an actual observation both $P_{in}$ and $\alpha_{in}$ are
a-priori unknown, which suggests that reconstructing the 
polarimeter modulation efficiency, or the incident polarization 
$P_{in}$ is subject to additional uncertainty.
In many practical cases this is not a limitation for 
reasons that we now discuss. 

Figure~\ref{fig:fig_eff_Pin.ps} shows that in cases where the incident
polarization is known to be small the modulation efficiency is to a
good approximation constant that does not depend either on $P_{in}$ or
on $\alpha_{in}$ (for the simple case of a single HWP see 
Equations~\ref{eq:eps_single} and~\ref{eq:eps_single_smallP}).
The region where the approximation 'small incident polarization'
is valid depends on the construction parameters of the HWP and 
the detection bandwidth. For example, 
for the top middle and bottom left panels the approximation is 
valid for $P_{in} \simlt 0.2$. It is valid for a much larger range
of $P_{in}$ when using a 5-stack (see right panels), and even for
the three-stack when it is used reasonably close to the designed band center
(see middle bottom panel). 

We note that the HWPs that are used for the calculations 
shown in Figure~\ref{fig:fig_eff_Pin.ps} are each designed
for a center frequency of 300~GHz (see Table~\ref{tab:pars}).
Therefore the top left panel that shows the largest variation
of the modulation efficiency with incidence angle is never
likely to be used in practice. It describes a single HWP optimized
for 300~GHz that is used for a band around 150~GHz. This panel 
is only shown for didactic purposes. 

We find then that in many practical situations there is 
a unique relation between $P_{in}$ and $P_{out}$, a relation
that does not depend on the orientation angle of $\alpha_{in}$.
In the more general case when the value of $P_{out}$ depends 
both on $P_{in}$ and on $\alpha_{in}$, the value of $\alpha_{in}$
needs to be determined first from the IVA. This is 
straight forward for a single HWP because the phase $\phi$ of
the IVA is equal to $\alpha_{in}/2$ for any detection bandwidth;
see for example the bottom left panel of Figure~\ref{fig:me_bandwidth_alpha.ps}.
The case of an AHWP is discussed in Section~\ref{subsec:alphavphi}, 
but the conclusion is that for a specified bandwidth there is a
unique relationship between the phase $\phi$ and the angle $\alpha_{in}$.
Therefore the procedure for finding $P_{in}$ is to first determine
$\alpha_{in}$ using the phase of the IVA and then use the relation
between $P_{out}$ and $P_{in}$ that is appropriate for this $\alpha_{in}$.

Laboratory measurements of modulation efficiency typically use 
incident polarizations that are close to $P_{in} = 1$ in order to increase
the signal to noise ratio of the measurement. Figure~\ref{fig:fig_eff_Pin.ps}
demonstrates that determinations of $\epsilon$ depend on the 
polarization angle $\alpha_{in}$. An efficiency value that 
was determined in the laboratory using a particular angle 
$\alpha_{in}$ will not in general correspond to the 
modulation efficiency of the polarimeter during actual observations
for which $\alpha_{in}$ is not known. 
A simple remedy is to align the incident polarization in the laboratory 
such that $\alpha_{in} = 45$~degrees. For that particular value the 
efficiency $\epsilon_{45}$ is a constant as a function of $P_{in}$ and is equal 
to the same efficiency that would be measured with small incident 
polarizations. Table~\ref{tab:Pout_Pin} summarizes this point
in a quantitative way. The values shown give the efficiency 
expected with $\alpha_{in}=45$~degrees for different frequency
bands and for different HWP configurations. The upper and lower
values marked with $\pm$ give the additional increments of efficiency 
that would be determined if $\alpha_{in}$ was 90 (for plus) or 0 (for minus) 
degrees. For example, the modulation efficiency of a single HWP 
(that is constructed according to the parameters given in Table~\ref{tab:pars})
at 150~GHz with a bandwidth of $\pm$30~GHz is 0.5 when measured 
with $\alpha_{in}=45$~degrees. This value of $\epsilon$ does not depend on the 
magnitude of $P_{in}$. Yet for observations with $P_{in}=0.1$ 
the modulation efficiency would be 0.055 (0.045) for  
$\alpha_{in}=90\, (0)$~degrees. Whereas for the 
single HWP the variation in modulation efficiency could be as large
as 10\%, it is about 2\% or smaller with the three- or five-stack. 
These values depend on the construction parameters of the HWP and 
on the detection bandwidth and thus can not be taken as general. 

\begin{figure}[t]
\centering
\includegraphics[width=10.5cm, angle=270]{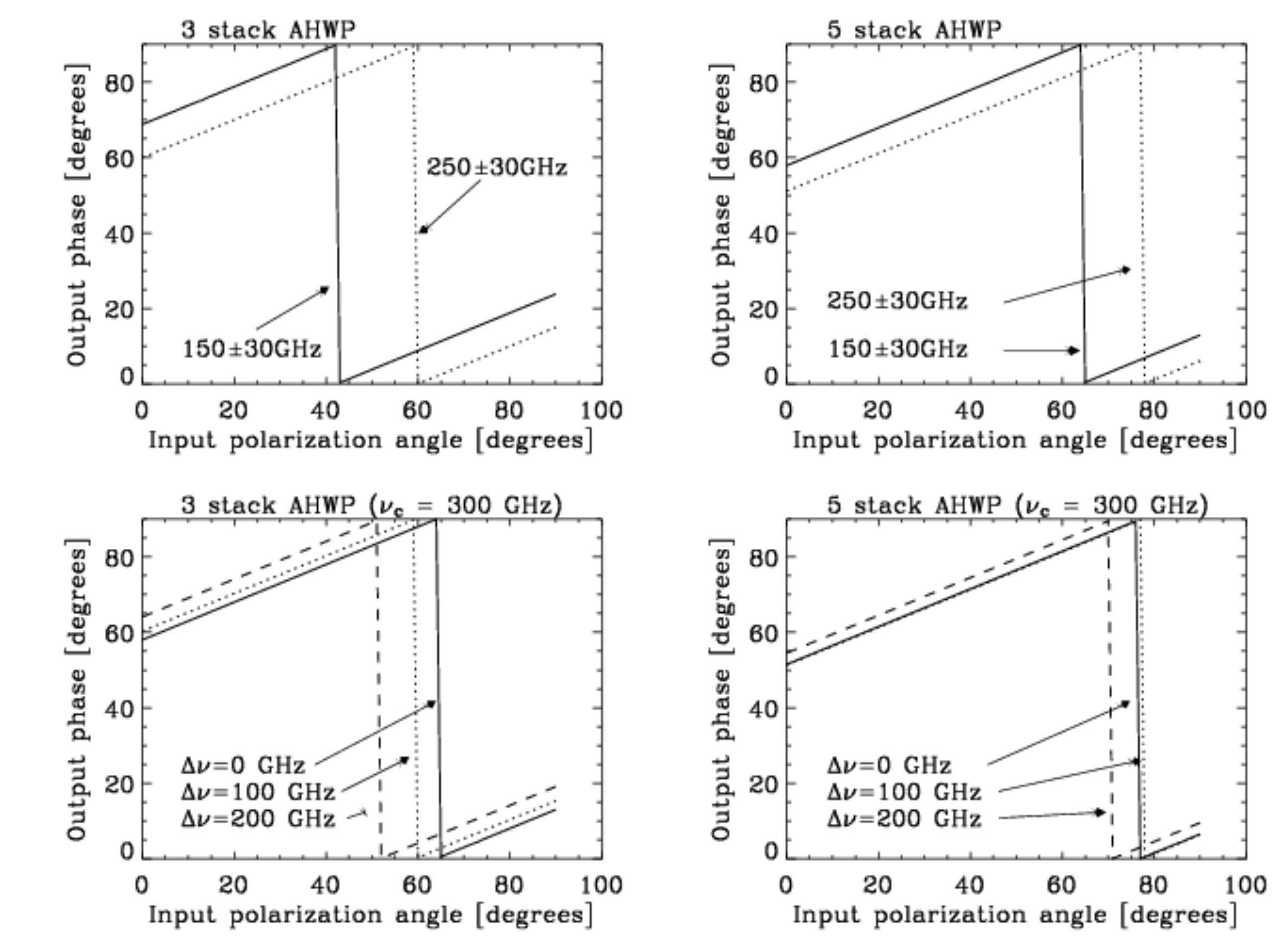}\vspace{-1.cm}
\caption{\footnotesize \setlength{\baselineskip}{0.95\baselineskip} 
	The output phase angle of the three- (left) and the five-stack
	(right) AHWPs as a function of the input
	polarization angle. The top panels give results
	for 150 (solid) and 250~GHz (dot), each with a fixed bandwidth 
	of $\pm$~30~GHz.
	The bottom panels give results for a fixed center 
	frequency of 300~GHz with bandwidths of $\pm$~0~(solid),
	$\pm$100~(dot), and $\pm$200~(dash)~GHz.}
\label{fig:phase_alpha}
\end{figure}

\subsection{$\phi$ vs. $\alpha_{in}$}
\label{subsec:alphavphi}

In Section~\ref{sec:pout_vs_pin} we investigated how the measured
degree of polarization $P_{out}$ relates to the input polarization $P_{in}$. 
We now quantify a similar relationship between $\phi$ and
$\alpha_{in}$. Figure~\ref{fig:phase_alpha} shows the phase angle as a
function of the input polarization angle for the three- and five-stack
AHWPs. (Recall that $\phi$ is a constant over frequency 
for a single HWP, see the bottom left panel of 
Figure~\ref{fig:me_bandwidth_alpha.ps}.) 
The panels show that $\alpha_{in}$ and $\phi$ have a linear
relationship with a slope of 0.5, and that this slope does not depend
on the construction parameters of the HWP nor on the detection
bandwidth. However, the phase offset {\it is} a function of 
$\nu_c$ and $\Delta \nu$. 

The magnitude of the phase offset $\phi_{0}$ is a critical parameter in the
reconstruction of an unknown incident polarization angle $\alpha_{in}$. Since this phase 
offset is a function of the spectral response of the instrument it can either be 
calculated, if the spectral response is known, or measured in the 
laboratory by varying $\alpha_{in}$ of a known source and extracting  $\phi$ 
from the IVA. 
(See however Section~\ref{sec:spectrum} for important caveats.)
Errors in this calibration will propagate to errors in the determination 
of  $\alpha_{in}$ for a source whose polarization properties are not known.

The direction of rotation of the AHWP affects the relationship between
$\phi$ and $\alpha_{in}$.  With single HWP $\phi = \pm
\alpha_{in}/2$, where the sign is determined by the 
direction of rotation. For our particular choice of directions (see
Figure~\ref{fig:setup}) we have $\phi = +\alpha_{in}/2$.  However, 
the orientation angles of the stack of plates 
break the rotational symmetry for an AHWP and in general there are four 
possible choices 
\beqna
\label{eq:ccw}
\phi &=& \pm \frac{1}{2} \alpha_{in} + \phi_0, \\
\label{eq:cw}
\phi &=& \pm \frac{1}{2} \alpha_{in} + \frac{\pi}{2} - \phi_{0}.
\eeqna
In Equation~\ref{eq:cw} the phase offset is no longer $\phi_0$, but $\frac{\pi}{2} - \phi_{0}$.
For our particular choice, where both $\rho$ and the AHWP
orientation angles $\vec{\theta}$ are counterclockwise in the $xy$
plane (as shown in Figure~\ref{fig:setup}), 
Equation~\ref{eq:ccw} with a plus sign gives the relevant
functional dependence.

\begin{table}[t]
\begin{center}
	\begin{tabular}[h]{|c|c|c|c|c|} \hline
                                         & CMB            & Dust           & Lab         \\ \hline
150~$\pm$~30~GHz  &   $\phi_0 =57.86$    & $56.69$     & $57.33$     \\ \hline
250~$\pm$~30~GHz   &   $51.12$   &$51.16$    & $51.14$      \\ \hline
420~$\pm$~30~GHz  &   $53.85$    &$54.50$    & $54.49$    \\ \hline
	\end{tabular}
\end{center}
\begin{center}
	\begin{tabular}[h]{|c|c|c|c|} \hline
                                         & CMB $-$ Dust  & CMB $-$ Lab & Dust $-$ Lab  \\ \hline
150~$\pm$~30~GHz  &   $\Delta \phi = 1.17~( \Delta\alpha_{in}=2.34)$       & $0.53~( 1.06)$       & $-0.64~( -1.28)$            \\ \hline
250~$\pm$~30~GHz   &   $-0.04~( -0.08)$    & $-0.02~( -0.04)$    & $0.02~( 0.04)$          \\ \hline
420~$\pm$~30~GHz  &   $-0.65~( -1.3)$    & $-0.64~( -1.28)$    & $0.01~( 0.02)$         \\ \hline
	\end{tabular}\vspace{-0.8cm}
\caption{\footnotesize \setlength{\baselineskip}{0.95\baselineskip}
	Top: The offset angles with four different spectra are shown.
	Bottom: The difference of the offset phase between different spectra. 
	The number in a parenthesis is the difference in terms of the polarization angle $\alpha_{in}$ on the sky.
	A unit of the phase is in degrees.}
\label{tab:incidentspectrum}
\end{center}
\end{table}

\section{Spectrum of Incident Radiation}
\label{sec:spectrum}

So far we assumed an incident radiation spectrum that was constant
with frequency.  We now address the more general case where the
spectrum of the incident radiation is a function of frequency. In this
case the phase offset $\phi_{0}$ depends on the details of this
spectrum. Since $\phi_{0}$ is required for reconstruction of $P_{in}$
and $\alpha_{in}$, the consequence is that knowledge of the incident
spectrum is also required.

To assess this effect quantitatively we consider three distinct spectra: 
(1) black body with the temperature of the  
cosmic microwave background radiation, (2) black body with a temperature 
of 300~K, and (3) galactic dust. We choose these spectra
because they are relevant for calibration and for measurements of the polarization 
of the CMB at frequencies between 100 and 500~GHz.
We assume the following spectra $I(\nu)$,
\beqna
I_{CMB}(\nu) &=&  B(T_{CMB}, \nu),\\ 
I_{dust}(\nu) &=& A \nu^\gamma B(T_{dust}, \nu), \\
I_{lab}(\nu) &=& B(T_{lab}, \nu),
\eeqna
\beq
B(T, \nu) = \frac{2\pi h}{c^2} \frac{\nu^3}{\mbox{e}^{\frac{h\nu}{k_B T}} - 1},
\eeq
where $B$ denotes a black body spectrum, $T_{CMB}=2.73$~K, $A=4\times10^{-7}$, $\gamma=1.75$,
$T_{dust}=18$~K, and $T_{lab}=300$~K. We assume that
the fractional polarization and the polarization angle of the incident
radiation do not depend on frequency.
We also assume that the degree of linear polarization of the CMB, 
of the galactic dust, and of a 300 K black body source are
$P_{CMB} = 1\times10^{-6}$, $P_{dust} = 0.1$, and $P_{lab} = 1$, respectively.

The calculated phase offsets of a five-stack AHWP are summarized in
Table~\ref{tab:incidentspectrum}.  The top table shows the phase
offset in unit of degrees. The bottom table shows the difference of
the phase offsets between the different spectra. The parentheses
indicate the level of difference in terms of polarization angle
$\alpha_{in}$.

Assume that a 300~K source is used in the laboratory to calibrate the
phase offset and that the laboratory measurement agrees with the phase
offsets given in the right hand column of the top part of
Table~\ref{tab:incidentspectrum}. If these values are used for either
CMB or dust observations, they would give rise to errors in position
angle of the polarization on the sky as given in parentheses in the
two right columns of the bottom table. The correct prescription is to
validate the design of the instrument using the laboratory
measurements and then use the predicted phase offsets given
assumptions or measurements of the spectra of the sources. An
uncertainty in the knowledge of the spectrum would give an uncertainty
in the determination of $\alpha_{in}$.  The designer of a polarimeter
with an AHWP should plan for this uncertainty and its mitigation
during the analysis of the data.

\begin{figure}[tphb]
\centering
\includegraphics[width=9.cm, angle=270]{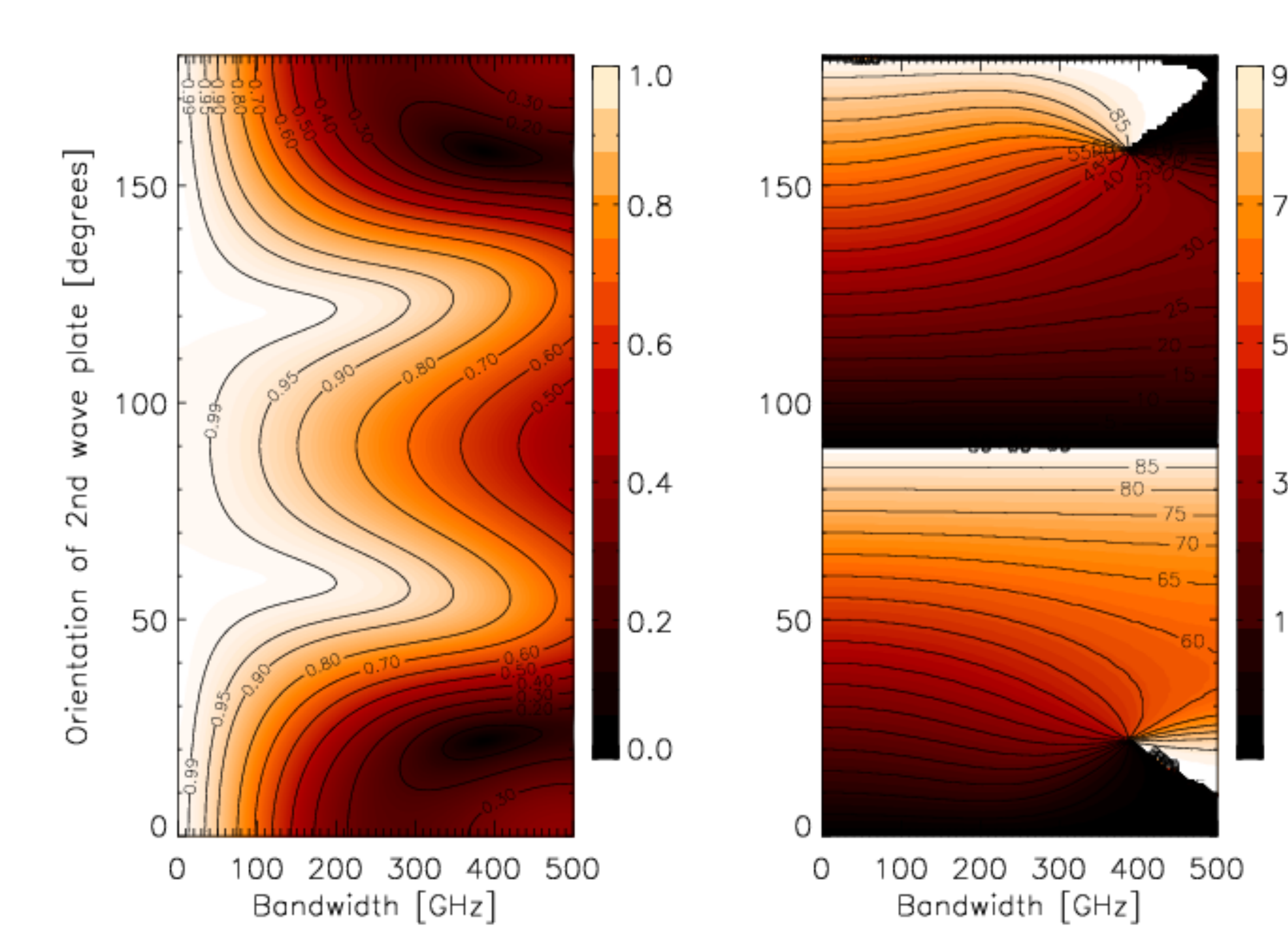}
\vspace{-1cm}
\caption{\footnotesize \setlength{\baselineskip}{0.95\baselineskip}
	The modulation efficiency (left) and the phase offset 
	(right) of the three-stack AHWP as a function of
	the angle of the second plate $\theta_2$ and the bandwidth $\Delta
	\nu$ around a center frequency of $\nu_{WP}$ = 300~GHz. 
      The color scale of the phase offset is in units of degrees.  In both
	plots, the input polarization angle is $\alpha_{in}=0$.}
\label{fig:3ahwp_twocont.ps} 
\end{figure}

\begin{figure}[tphb]
\centering
\includegraphics[width=9.cm, angle=270]{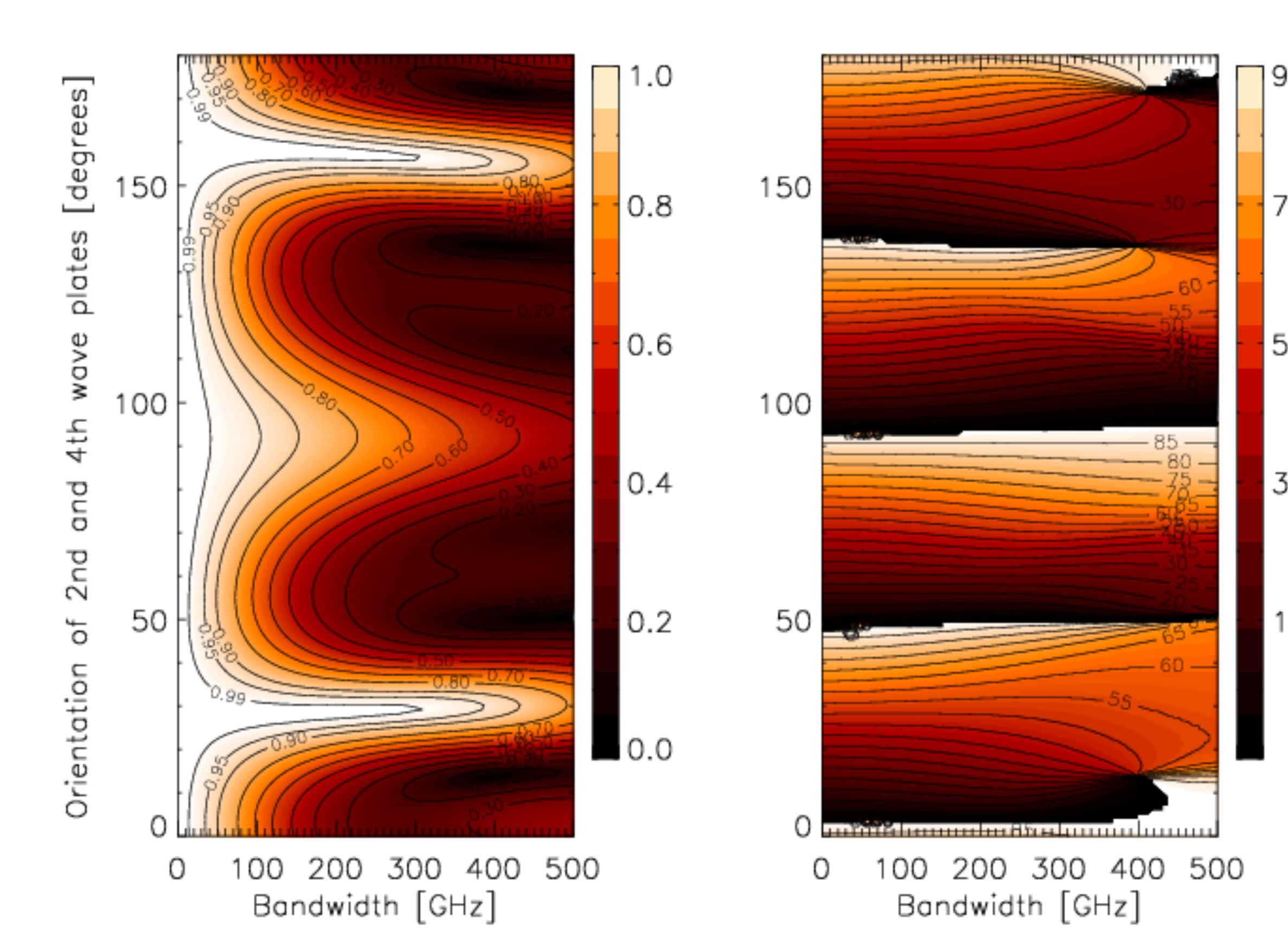}
\vspace{-1cm}
\caption{\footnotesize \setlength{\baselineskip}{0.95\baselineskip} 
	The modulation efficiency (left) and the phase offset (right)
	of the five-stack AHWP as a function of the orientation
	angles of the second and fourth plates. The other angles are fixed at the values
      given in Table~\ref{tab:pars}.}
\label{fig:5ahwp_24.ps}
\end{figure}

\begin{figure}[tphb]
\centering
\includegraphics[width=9.cm, angle=270]{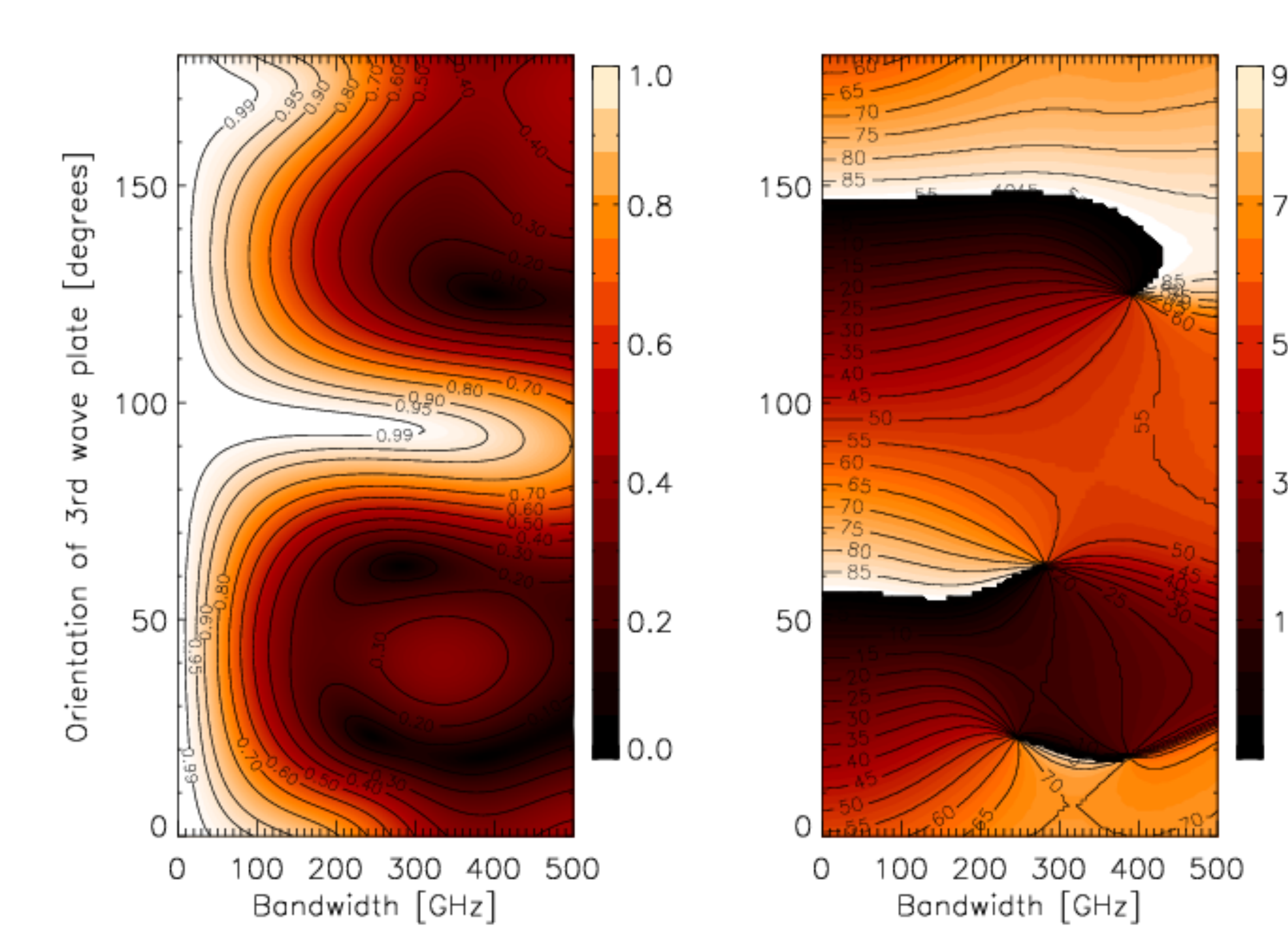}
\vspace{-1cm}
\caption{\footnotesize \setlength{\baselineskip}{0.95\baselineskip} 
	The modulation efficiency (left) and the phase offset (right)
	of the five-stack AHWP as a function of the orientation
	angle of the third plate. The other angles are fixed at the 
      values given in Table~\ref{tab:pars}.}
\label{fig:5ahwp_3rd.ps}
\end{figure}

\begin{figure}[tphb]
\centering
\includegraphics[ width=9.cm, angle=270]{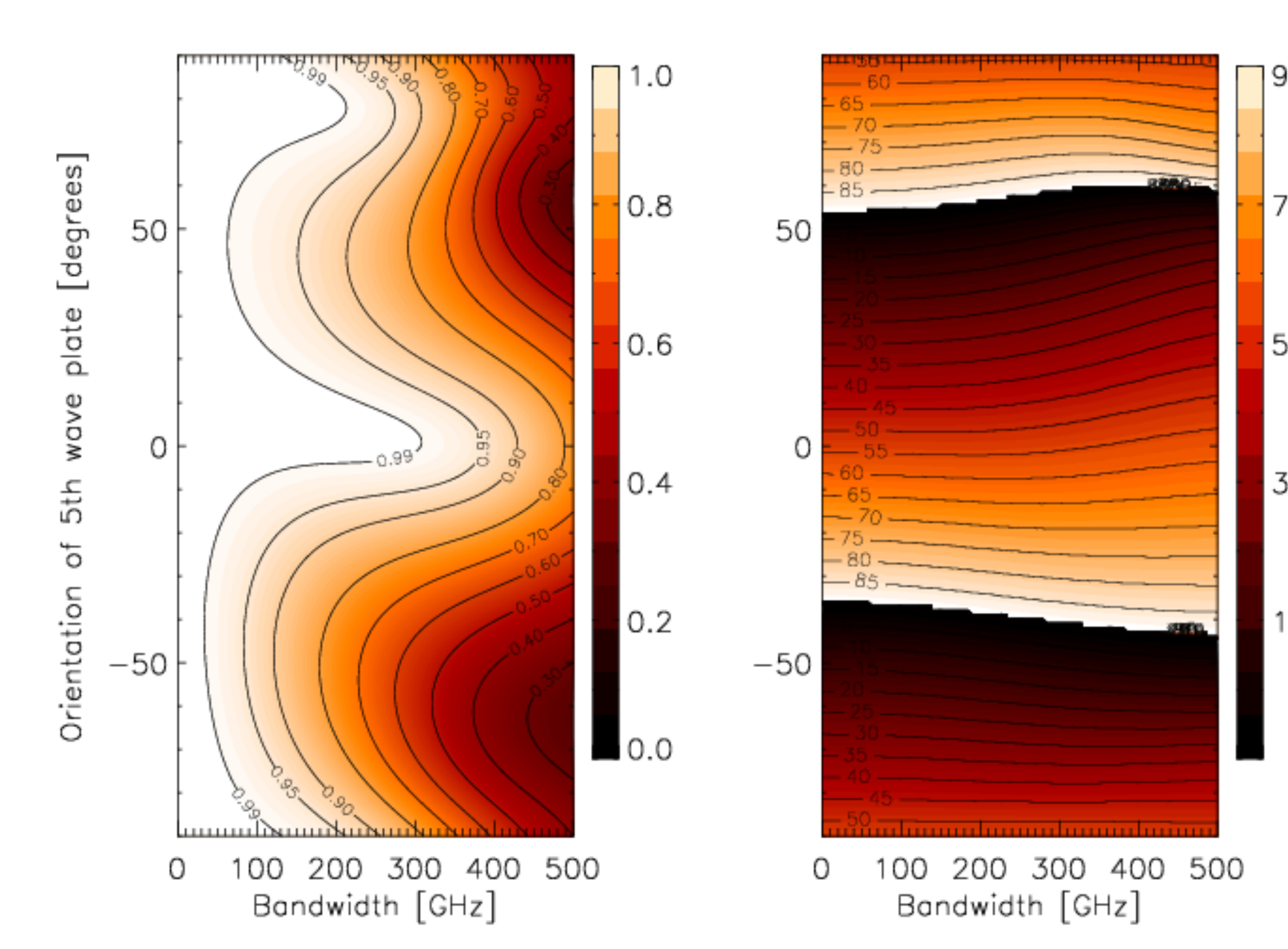}
\vspace{-1cm}
\caption{\footnotesize \setlength{\baselineskip}{0.95\baselineskip} 
	The modulation efficiency (left) and the phase offset  (right) of the 
      five-stack AHWP as a function of the orientation angle of the fifth plate.
      The other angles are fixed at the values given in Table~\ref{tab:pars}.}
\label{fig:5ahwp_5th.ps}
\end{figure}

\begin{figure}[tphb]
\centering
\includegraphics[ width=9.cm, angle=270]{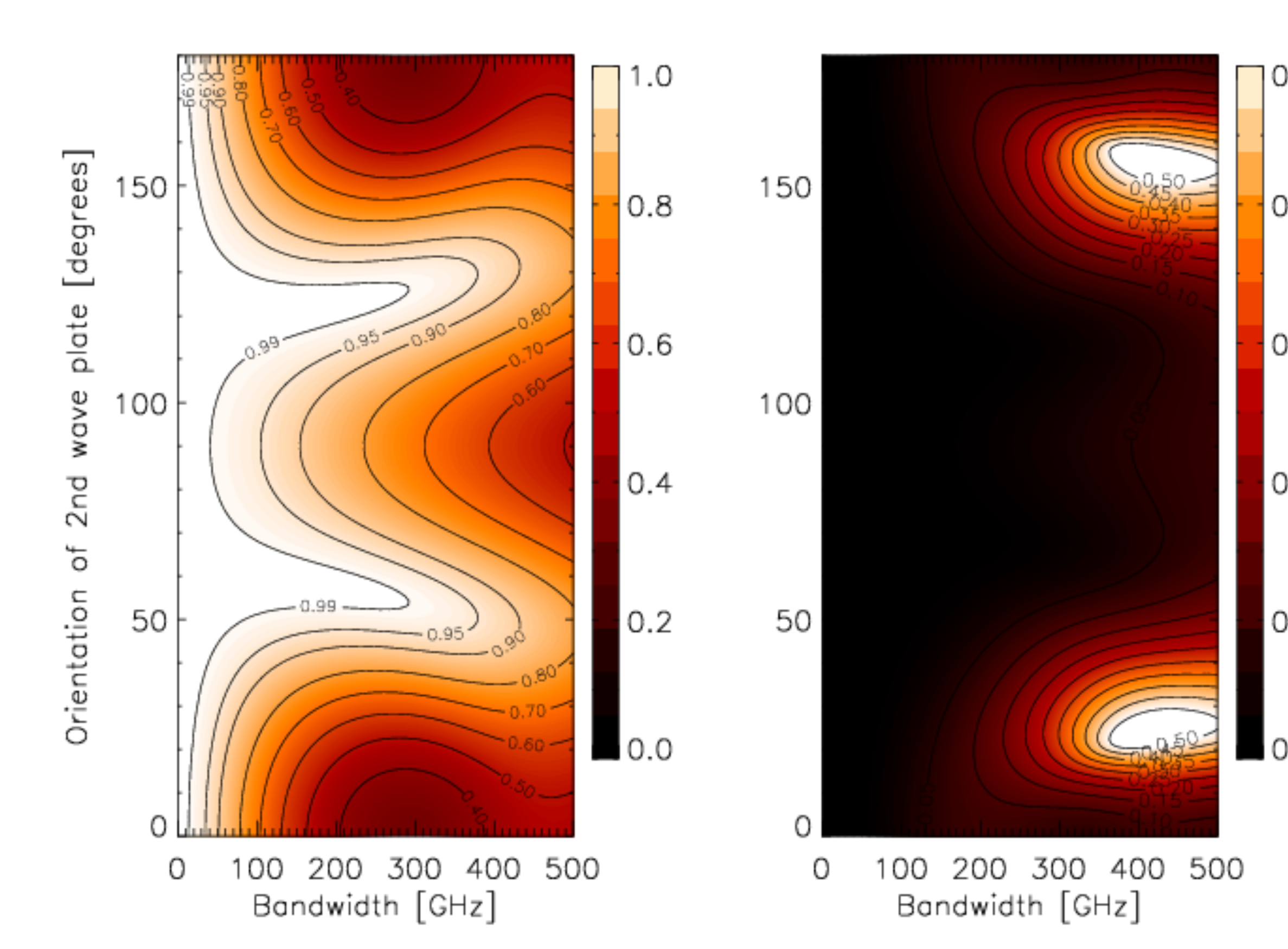}
\vspace{-1cm}
\caption{\footnotesize \setlength{\baselineskip}{0.95\baselineskip} 
	The modulation efficiency of a three-stack AHWP
      based on Equation~\ref{eq:Pout_hananyetal} (left) 
      and the difference between this efficiency and the one calculated in 
      Figure.~\ref{fig:out_3AHWP.180.comp.diff.ps} (right).
      }
\label{fig:out_3AHWP.180.comp.diff.ps}
\end{figure}

\section{AHWP Performance vs. orientation angles $\vec{\theta}$}
\label{sec:performance.vs.theta}

We have already pointed out in an earlier publication that it is
relatively easy to achieve a high modulation efficiency with the
three and five-stacks in terms of the requirements on the relative
orientation of the plates~\cite{hanany05}. In this section we expand
on our earlier work and give a more thorough discussion.
All of the analysis in this section assume a constant spectrum 
for the incident radiation.

The left panels of
Figures~\ref{fig:3ahwp_twocont.ps}, \ref{fig:5ahwp_24.ps},
\ref{fig:5ahwp_3rd.ps}, and \ref{fig:5ahwp_5th.ps} 
give contour plots for the modulation efficiency as a function of the orientation
of the plates in the stacks. The modulation efficiency is calculated based on
Equation~\ref{eq:Pout} with Equations~\ref{eq:Maluslaw_freqsum} and \ref{eq:me}. 
The right panels of the same figures show the
phase offset as defined in Equation~\ref{eq:IVA}.

For a three-stack Title~\cite{title81} showed that the highest modulation
efficiency is achieved with a set of angle $\vec{\theta} =
(0,58,0)$. The left panel of Figure~\ref{fig:3ahwp_twocont.ps} shows
that this modulation is a weak function of the orientation of the
middle plate near peak efficiency.  The right panel of Figure~\ref{fig:3ahwp_twocont.ps}
shows that for a second plate orientation  $80 \simlt \theta_{2} \simlt 100$~degrees
the phase offset $\phi_{0}$ is essentially independent of detection
bandwidth. This orientation angle, however, does not give the broadest
range of frequencies for high modulation efficiency.  On the other
hand, with an angle of 58 degrees, which gives the broadest range of
modulation efficiency, the phase offset has stronger dependence on the
detection bandwidth. Experiment designers need to consider this
trade-off between bandwidth for high modulation and for constant phase
offset.

The three-stack has zero modulation efficiency at $\theta_2$ close to
20 and 160~degrees and detection bandwidth of 400~GHz.  This is because there
is a strong variation of the phase offset angle $\phi_{0}$ with
frequency near these parameters.
Therefore, these points in the parameter space
give the resultant IVA zero modulation amplitude, and correspondingly
no phase can be defined as demonstrated by the singularities in
the phase offset panel.
The color discontinuity extending from the
phase offset singularity toward bottom right is a consequence of phase
offset periodicity. It is neither an artifact nor a real
discontinuity. Phases that are larger than 90~degrees are
interpreted as positive values close to zero. Similar features appear
in Figures~\ref{fig:5ahwp_24.ps}, \ref{fig:5ahwp_3rd.ps}, and
\ref{fig:5ahwp_5th.ps}.

With the five-stack, achieving high modulation efficiency requires
higher accuracy of alignment of the second, third, and
fourth wave plates than that required in the case of the three-stack. 
Little accuracy is required from the orientation of the fifth plate 
in the five-stack. The efficiency is most sensitive
to the orientation of the second and fourth plates, and an accuracy of 5~degrees
is required to maintain efficiency higher than 0.95 over 300$\pm$150~GHz.

In a previous publication~\cite{hanany05} we gave results 
for the modulation efficiency that was based on the following expression 
\begin{equation}
\label{eq:Pout_hananyetal}
P_{out} = \langle \frac{ I_{max} - I_{min}}{I_{max} + I _{min}} \rangle, 
\end{equation}
which is different from the more correct definition given in 
Equation~\ref{eq:Pout}.
The left panel of Figure~\ref{fig:out_3AHWP.180.comp.diff.ps} is the modulation efficiency based
on Equation~\ref{eq:Pout_hananyetal} with the same parameters 
that produced Figure~\ref{fig:3ahwp_twocont.ps}. 
The right panel shows the differences between the two results.
The modulation efficiency in Figure~\ref{fig:3ahwp_twocont.ps} accounts for the 
phase variation of the IVA curves as a function of frequency.
In contrast the modulation efficiency in Figure~\ref{fig:out_3AHWP.180.comp.diff.ps} does
not encode this variation.

\section{Discussion}
\label{sec:discussion}

We analyzed the performance of a three- and five-stack AHWP
polarimeters operating in the sub-millimeter wave band. 
Let us summarize the points that have been discussed and
make some additional comments where appropriate.
\begin{itemize}
\item Three- and five-stack AHWP polarimeters provide 
broad bandwidth with high modulation efficiency. 
\item Their IVA has a phase offset that depends on the 
construction parameters of the stack, on the spectral 
response of the instrument, and on the spectrum of incident radiation. 
(our discussion assumed that the degree of polarization 
and the angle were independent of frequency with the detection bandwidth.)
\item If the spectral response of the instrument, 
and the spectrum of the source are known, 
then measurements of the phase of the IVA can 
give the orientation angle of the incident polarization. 

We note that in many cases much of the radiation incident on 
the detector is due to emission by the telescope itself. 
If this emission is polarized it too will affect the phase of 
the IVA and hence the measurement of the angle of incident polarization.
\item measurement uncertainties in either the spectral response 
of the instrument or the spectrum of the source translate 
to uncertainties in the reconstruction of the angle of incident polarization. 
The amount of uncertainty needs to be assessed on a case-by-case basis. 

\item Measurements of the amplitude of the IVA, 
which gives the degree of output polarization, 
can be uniquely inverted, in most cases, to give 
the input polarization if the modulation efficiency is known. 
\item In some cases information about the angle of 
the incident polarization needs to be used together 
with the modulation efficiency to find the incident degree of polarization. 
\item Laboratory measurement to find the modulation efficiency 
that are conducted with a source that has high degree of polarization 
should have an incident polarization angle of 45 degrees. 
At this angle the measured efficiency is the same as 
would be measured at any angle when $P_{in}$ is small. 
\item We discussed how the modulation efficiency 
and phase offset of the polarimeters depend on errors 
in the orientation of the plates. Generally, an accuracy 
of few degrees is sufficient to ensure close the ideal performance. 
\item We discussed the how the incident spectrum of the radiation 
affects the IVA and the extraction of the parameters of the incident radiation. 

Our analysis assumed a spectral response of the instrument that was top-hat in
shape over a range in frequencies. This is an idealization. In any practical 
instrument, the entire spectral response of the instrument is necessary in 
order to reconstruct the parameters of the incident polarization. 

\end{itemize}


\newpage

\section*{List of Figure Captions}

Fig. 1. A schematic diagram of the HWP polarimeter model. The transmission axis of a linear polarizer is parallel to the $x$ axis. 

Fig. 2. IVA for monochromatic light (top panels) and for broadband radiation (bottom panels) for a single HWP, a three-stack AHWP, and a five-stack AHWP (left to right). See Table~\ref{tab:pars} for the parameters of the plates and for the details about the simulations used for the calculations. Frequencies of 150~(solid), 200~(dash), 250~(dot), 300~(dash-dot)~GHz are used for the case of monochromatic light. For the broadband case we use $150\pm30$~GHz~(solid) and $250\pm30$~GHz~(dot). In all the panels, the maximum intensity is normalized to 1.

Fig. 3. Modulation efficiency $\epsilon = \epsilon(\nu, \Delta\nu=0, \alpha_{in}=0,\vec{\theta})$ (top) and the phase offset $\phi_0 = \phi(\alpha_{in}=0, \nu, \Delta\nu=0, \vec{\theta})$ (bottom) for the single HWP (left) and the three- (middle) and the five-stack (right)  as a function of frequency.  

Fig. 4. Top: Modulation efficiency of the single HWP, the three- and the five-stack AHWPs  as a function of detection bandwidth for input polarization angle of 0 (solid line), 22.5 (dot), 45 (dash), 67.5 (dot-dash), and 90 (three-dot dash) degrees. Bottom: Output phase angle of the single, three-, and five-stack as a function of detection bandwidth for the same input polarization angles as the top panels. For both the modulation efficiency and the phase, $\nu_c = \nu_{WP}$.

Fig. 5. The extracted degree of polarization $P_{out}$ as a function of the degree of polarization of the incident light $P_{in}$ for the single-, three-, and five-stack. Each curve corresponds to the input polarization angle of 0 (solid line), 22.5 (dot), 45 (dash), 67.5 (dot-dash), and 90 (three-dot dash) degrees. The frequency and the bandwidth are $\nu_c\pm\Delta\nu = 150\pm30$~GHz (top) and $250\pm30$~GHz (bottom). For all the panels, $\nu_{WP} = 300$~GHz.

Fig. 6. The output phase angle of the three- (left) and the five-stack (right) AHWPs as a function of the input polarization angle. The top panels give results	for 150 (solid) and 250~GHz (dot), each with a fixed bandwidth of $\pm$~30~GHz.	The bottom panels give results for a fixed center frequency of 300~GHz with bandwidths of $\pm$~0~(solid), $\pm$100~(dot), and $\pm$200~(dash)~GHz.

Fig. 7. The modulation efficiency (left) and the phase offset (right) of the three-stack AHWP as a function of the angle of the second plate $\theta_2$ and the bandwidth $\Delta \nu$ around a center frequency of $\nu_{WP}$ = 300~GHz. The color scale of the phase offset is in units of degrees. In both plots, the input polarization angle is $\alpha_{in}=0$.

Fig. 8. The modulation efficiency (left) and the phase offset (right) of the five-stack AHWP as a function of the orientation angles of the second and fourth plates. The other angles are fixed at the values given in Table~\ref{tab:pars}.

Fig. 9. The modulation efficiency (left) and the phase offset (right) of the five-stack AHWP as a function of the orientation angle of the third plate. The other angles are fixed at the values given in Table~\ref{tab:pars}.

Fig. 10. The modulation efficiency (left) and the phase offset  (right) of the five-stack AHWP as a function of the orientation angle of the fifth plate. The other angles are fixed at the values given in Table~\ref{tab:pars}.

Fig. 11. The modulation efficiency of a three-stack AHWP based on Equation~\ref{eq:Pout_hananyetal} (left) and the difference between this efficiency and the one calculated in Figure.~\ref{fig:out_3AHWP.180.comp.diff.ps} (right).

\section*{Table Caption}
Tab. 1. Parameters of the wave plates and parameters used in the simulations to calculate the IVA.

Tab. 2. The modulation efficiency at $P_{in} = 0.1$ with $\alpha_{in}= 45$~degrees is shown. The modulation efficiency is calculated as a slope of $P_{out}-P_{in}$ relationship in Figure~\ref{fig:fig_eff_Pin.ps}. The quoted errors are $\epsilon_{max} - \epsilon_{45}$ and $\epsilon_{min} - \epsilon_{45}$, where $\epsilon_{45}$ corresponds to the modulation efficiency at $\alpha_{in} = 45$~degrees at $P_{in} = 0.1$. The maximum and the minimum modulation efficiency corresponds to $\alpha_{in} = 90$ and $0$~degrees, respectively.

Tab. 3. Top: The offset angles with four different spectra are shown. Bottom: The difference of the offset phase between different spectra. The number in a parenthesis is the difference in terms of the polarization angle $\alpha_{in}$ on the sky. A unit of the phase is in degrees.

\end{document}